\documentclass[aps,apl,twocolumn,showpacs,superscriptaddress]{revtex4}

\usepackage{graphicx} % Include figure files

\begin{document}

\title
{Silicon and III-V compound nanotubes: structural and electronic
properties}

\affiliation{Department of Physics, Bilkent University, Ankara
06800, Turkey}
\author{E. Durgun}
\author{S. Tongay}
\author{S. Ciraci}
\email{ciraci@fen.bilkent.edu.tr}

\date{October 22, 2004}

\begin{abstract}
 Unusual physical properties of single-wall carbon nanotubes have
 started a search for similar tubular structures of
 other elements. In this paper, we present a theoretical analysis of single-wall
 nanotubes of silicon and group III-V compounds. Starting from precursor graphene-like structures
 we investigated the stability, energetics and electronic
 structure of zigzag and armchair tubes using first-principles pseudopotential plane wave
 method and finite temperature \emph{ab-initio} molecular dynamics
 calculations. We showed that (\emph{n},0) zigzag and (\emph{n,n}) armchair nanotubes of silicon having $n \geq 6$ are
stable but those with $n < 6$ can be stabilized by internal or
external adsorption of transition metal elements.  Some of these
tubes have magnetic ground state leading to spintronic properties.
We also examined the stability of nanotubes under radial and axial
deformation. Owing to the weakness of radial restoring force,
stable Si nanotubes are radially soft. Undeformed zigzag nanotubes
are found to be metallic for $6 \leq n \leq 11$ due to curvature
effect; but a
 gap starts to open for $n \geq 12$. Furthermore, we identified stable tubular structures formed by
stacking of Si polygons.
 We found  AlP, GaAs, and GaN (8,0) single-wall nanotubes stable and
 semiconducting. Our results are compared with those of
 single-wall carbon nanotubes.

\end{abstract}


\maketitle

\section{introduction}
Carbon nanotubes are unique one dimensional
nanostructures\cite{iijima} with their exceptional mechanical,
electronic and magnetic
properties.\cite{dress,saito,mintmire,review} While the use of
single-wall carbon nanotubes (SWCNT) requires a completely new
paradigms in the development of nanodevices, Si still continues to
attract interest for electronic applications in nanotechnology.
Therefore, Si based nanotubes have been subject of experimental
and theoretical analysis.

Even if a single-wall Si nanotube (SWSiNT) has never been
observed, theoretical predictions have been performed for various
kinds of Si tubes. Fagan \textit{et al.}\cite{Fagan1,Fagan2} have
investigated the structural and electronic properties of chiral
SWSiNTs based on Density Functional Theory (DFT). Barnard
\textit{et al.}\cite{Banard} have examined the dependence of heat
of formation and binding energy of SWSiNTs on their radius and
chirality. The stability of (10,0) SWSiNT has been examined by
using empirical Monte-Carlo molecular dynamics method and found
that it is stable at finite temperature.\cite{Fagan2} Ivanovskaya
\textit{et al.}\cite{ivan} investigated hypothetical Si nanotubes
containing regular chains of metallocarbonedrenes using one
dimensional tight-binding model within H\"{u}ckel approximation.
By using \textit{ab-initio} calculations, Dumitrica \textit{et
al.}\cite{dumitrica} described how the smallest (2,2) and (3,0)
SiNTs are stabilized by the axially placed metal atoms from
different groups of periodic table. Ponomarenko \textit{et
al.}\cite{ponomarenko} studied the energetics and relative
stability of infinite and finite, clean and hydrogenated
open-ended Si nanotubes by using the extended Brenner potential.
The existence of H-doped stable tube-shaped finite SiNTs have been
predicted\cite{MZhang} and their electronic structures have been
compared with those of carbon nanotubes.\cite{RQZhang} Seifert
\textit{et al.}\cite{Seifert} have argued that structures of
silicate and SiH nanotubes are more stable than bare Si nanotubes.
Singh \textit{et al.}\cite{Singh} have investigated the stability
of finite and infinite hexagonal prismatic structures of Si with
$3d$ magnetic elements and predicted that such structures can be
stabilized through doping by the elements of transition metal
(TM). Fullerene-structured Si tubulars, possibly based on
Si$_{24}$ have been produced.\cite{bjorn} More recently, the
successful synthesis of multiwalled Si nanotubes has been
reported.\cite{YangSha} Now, SWSiNTs are no longer hypothetical
structures and it is not unrealistic to expect their fabrication
with controllable size and diameter. Similarly, achievements of
synthesis of single-wall BN nanotubes\cite{Loiseau} and GaN
\cite{weber,goldberger}, AlN \cite{tondare} thick-wall tubular
forms has increased the interest in the theoretical analysis of
compound nanotubes.\cite{cohen,smlee,cote,zhao} In addition, the
synthesis of Mo and W chalcogenid
nanotubes\cite{rapoport,feldman,tenne}, and also NiCl tubular and
cage structures have been realized.\cite{hacohen}

In this paper we present a theoretical analysis of Si- and III-V
compound nanotubes based on a state-of-the-art first-principle
calculations. Our work is concentrated mainly on the tube
structures which can be viewed as the rolling of graphene like
honeycomb planes of Si or III-V elements on a cylinder of radius
$R$. Starting from the precursor graphene-like honeycomb
structures we investigated their stability, energetics and
electronic properties of these nanotubes. Since O, O$_2$, Si, Au
and H are critical elements for various processes on Si, we also
examined the adsorption of these atoms on SWSiNT. Finally, we
studied the stabilization of unstable, small-diameter SWSiNTs
through the internal and external adsorption of transition metal
elements. In addition, we found that tubular structures which are
generated by stacking of triangles, pentagons and hexagons of Si
are stable and metallic. The (8,0) zigzag tubes of AlP, GaN and
GaAs are stable and semiconducting. The results obtained from the
present study have been compared systematically with those of
SWCNT. The stable tube structure predicted in this study are hoped
to motivate experimental research aiming at the synthesis of
various tubular structures of group IV elements and III-V and
II-VI compounds

\section{method}

We have performed first-principles plane wave
calculations\cite{payne,vasp} within DFT\cite{kohn} using
ultra-soft pseudopotentials.\cite{vasp,vander} The exchange
correlation potential has been approximated by Generalized
Gradient Approximation. (GGA)\cite{gga} Structures incorporating
TM atoms have been calculated using spin-polarized GGA. For
partial occupancies we use Methfessel-Paxton smearing
method.\cite{methfessel} The width of smearing is chosen between
0.01-0.1 eV depending on the system. All structures have been
treated by supercell geometry using the periodic boundary
conditions. To prevent interactions between adjacent structures a
large spacing ($\sim 10 \AA$) has been taken. Convergence with
respect to the number of plane waves used in expanding Bloch
functions and \textbf{k}-points in sampling the Brillouin zone are
tested for the parent bulk crystals as well as tubular structures.
In the self-consistent potential and total energy calculations
Brillouin zone of nanotubes has been sampled by (1x1x19) mesh
points in \textbf{k}-space within Monkhorst-Pack
scheme.\cite{monk} Calculations of graphene and graphite
structures have been carried out using (19x19x1) and (8x8x6)
\textbf{k}-point samplings, respectively. A plane-wave basis set
with kinetic energy cutoff $200 eV \leq \hbar^2
|\textbf{k}+\textbf{G}|^2/2m \leq 330 eV$ has been used. All
atomic positions and lattice parameters are optimized by using
conjugate gradient method where total energy and atomic forces are
minimized. The convergence for energy is chosen as 10$^{-5}$ eV
between two ionic step, and the maximum force allowed on each
atoms is 0.05 eV/$\AA$.

It should be noted that DFT based methods provide reliable
predictions for the ground state properties , but band gaps are
usually underestimated. Hence the energy band structure obtained
from the single particle energy eigenvalues of Kohn-Sham equations
are only approximations to the real energy bands. Proper many-body
self-energy corrections can be made by using GW method.\cite{gw}
Recently, GW energies are compared with DFT-LDA results of $(n,0)$
SWCNTs which indicates shifts of valence and conduction bands and
considerable increase of LDA band gap from 0.2 eV to 0.6
eV.\cite{miyake} It is suggested that GW corrections are small for
large radius SWCNTs. Performing first-principles many-body Green`s
function calculations Spataru \emph{et al.}\cite{louie} showed
that the optical spectrum of both semiconducting and metallic
small-radius SWCNTs exhibit important excitonic effects due to
quasi-one dimensional nature. It is interesting to note that while
the band gaps of (9,0), (12,0) and (15,0) zigzag SWCNTs have been
measured by Scanning Tunneling Spectroscopy \cite{lieber} to be
80, 42 and 29 meV, respectively, same band gaps have been
predicted by GGA calculation \cite{gulseren} to be 93, 78 and 28
meV, respectively.

The stability of the structures we studied is the most crucial
aspect of our work, since it provides valuable information for the
synthesis of these materials in future. In this respect an
extensive analysis of stability has been carried out for various
nanotubes. First, we applied a radial deformation to certain
nanotubes and optimized their structures to see whether they relax
to their original, undeformed circular forms under zero external
force. Furthermore, we have performed, finite temperature
\emph{ab-initio} molecular dynamics calculations up to 1000 K
using Nos$\acute{e}$ thermostat\cite{nose} for 250 time steps (0.5
ps) to check whether the optimized structure will be affected from
random thermal motion of atoms or they maintain their tubular form
at high-temperature. We believe that if there were any kind of
structural instability it would be initiated and also enhanced
within these time-steps at high temperature.

\section{honeycomb structure of silicon and III-V compounds}

\begin{figure}
\includegraphics[scale=0.45]{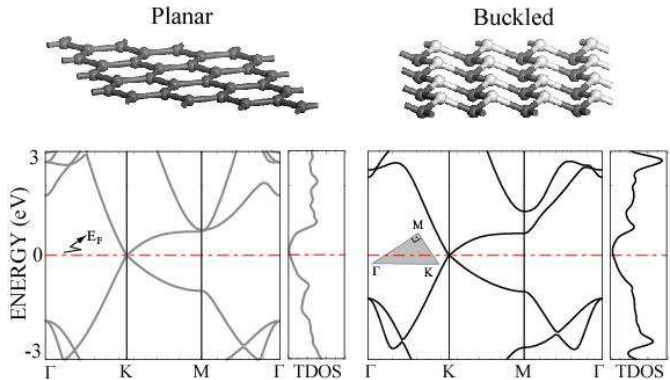} \caption{Band
structure and total density of states (TDOS) analysis for planar
and buckled honeycomb structure (HC) of Si. Light-lines correspond
to planar (having P6/MMM symmetry) and dark-lines correspond to
buckled structure (having P-3M1 symmetry). The zero of energy is
set to the Fermi level $E_F$. The inset shows the 1/12 of
hexagonal Brillouin zone.} \label{fig:graphene}
\end{figure}

One of the main difficulty for synthesizing Si nanotubes seems to
be the absence of 2D silicon layer similar to the graphene
structure of carbon. This is traced to the fact that in contrast
to carbon, $sp^3$-hybridization in Si is more stable than
$sp^2$-hybridization.\cite{parinello} In view of this situation we
examined whether the graphene-like 2D sheet of silicon can be
stable. Two dimensional hexagonal lattice forming a honeycomb
structure in the $xy$-plane has been periodically repeated along
$z$-axis with 10 $\AA$ spacing to minimize interlayer
interactions. In order to reduce the effects of the constraints to
be imposed by using the primitive unit cell we performed structure
optimizations on the (2x2) cell in the $xy$-plane. Our
calculations revealed that the planar structure (where all atoms
lie in the same plane) is metastable, but it is buckled by 0.45
$\AA$ relative vertical displacement of alternate atoms on the
hexagons. The gain of energy upon buckling is 30 meV/atom. The
binding energy is calculated to be 4.9 eV/atom which is 0.6 eV
lower than Si diamond structure  and the average distance between
nearest Si atoms is 2.2$\AA$. In the rest of the paper this
graphene-like structure will be specified as buckled honeycomb
structure. As shown in Fig.\ref{fig:graphene}, the detailed band
structure and total density of states (TDOS) analysis indicate
that both buckled and planar system have large band gaps along
$\Gamma$K and MK directions, but conduction and valence bands
cross the Fermi level at the \textbf{k}-point of Brillouin zone.
The electronic structure of the system does not change
significantly as a result of buckling, except some of the bands
split due to the lowering of the rotation symmetry. Using a
similar method but different pseudopotentials and exchange
correlation potential Takeda \emph{et al.}\cite{takeda} have
examined planar and buckled honeycomb structures of Si. Our
results obtained in four times larger cell hence allowing more
variational freedom are in overall agrement with the results in
Ref. \cite{takeda}. Moreover, we performed an \emph{ab-initio}
molecular dynamics calculations on 2x2 supercell providing further
evidence that buckled honeycomb structure is stable at 500K for
250 time steps.

Next we address whether a graphite like structure of Si (or
graphitic Si) can form? Our study distinguished chemisorption and
physisorption states in the interlayer interaction, in contrast to
only the physisorption state in graphite.\cite{ycwang} The
chemisorption state corresponding to a smaller lattice parameter
$c=6\AA$ is energetically more favorable, namely the binding is
5.1 eV which is 0.4 eV smaller than that of the bulk Si. We note,
however, that the lattice parameters, binding energies depend on
the approximation of exchange-correlation
potential.\cite{charlier,vdw} Present GGA calculations yield
relatively larger interlayer distance and requires the
incorporation of long range Van der Walls (VdW) interaction. Such
an analysis related with the binding of graphite showing that the
lattice parameter calculated by GGA is improved upon including VdW
attraction has been presented elsewhere.\cite{sefa}

Similar to Si, the honeycomb structures of AlP, GaAs and GaN are
found to be also stable, but less energetic relative to the bulk
crystal by 0.8, 1.1 and 0.6 eV per basis, respectively. However,
the buckling is not favored in order to hinder the formation of
dipole layer.

\section{Single wall silicon nanotubes}
\subsection{Energetics and Stability}
\begin{figure}
\includegraphics[scale=0.45]{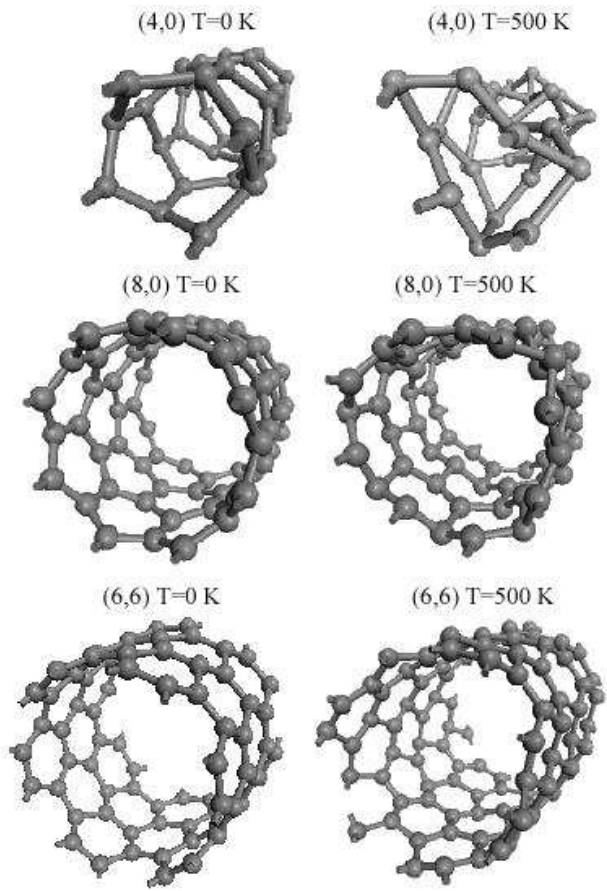} \caption{Structures of
(4,0), (8,0) and (6,6) SWSiNTs at T=0 and T=500K after 250 time
steps. Tubular structure has remained in (6,6) and (8,0) SWSiNT,
but (4,0) structure has clustered.} \label{fig:temp}
\end{figure}

Having discussed the stability of Si buckled honeycomb structure
(Si-HC), now we present our systematic analysis of $(n,0)$ zigzag
and $(n,n)$ armchair SWSiNTs for different $n$ values; namely
$n=3-14$ for zigzag and $n=3,6,9$ for armchair structures. The
(3,0) zigzag SWSiNT has clustered upon structure relaxation
indicating that it is not stable even at T=0 K. While the
structure optimization has resulted in a regular (4,0) and (5,0)
tubular structures, the $ab-initio$ MD calculations showed that
these nanotubes eventually transform into cluster at higher
temperatures as shown in Fig. \ref{fig:temp}. Significant
distortions can be easily noticed in (6,0) and also (7,0) SWSiNTs,
but tubular character and hexagonal structures on the surface have
remained. The (6,0) zigzag tube, which has a radius of $R=3.8 \AA$
as well as those with larger radii remain stable at temperatures
up to 800 K. Barnard \emph{et al.} \cite{Banard} also reported the
instability of (3,0) SWSiNT in their first-principles study, but
they considered (4,0) and (5,0) SWSiNTs as stable structures
depending on their geometry optimization performed at T=0 K.
Present results set a limit for fabricating small radius SWSiNTs.
The first and second nearest neighbor interactions between Si
atoms become relevant for the stability of small radius nanotubes
and causes clusterings, if $R < 3.8 \AA$. Similar behavior is also
obtained for $(n,n)$ armchair SWSiNTs. For example (3,3) SWSiNT is
clustered at 800 K in spite of the fact that geometry optimization
yields tubular structure at T=0 K. On the other hand, the (6,6)
tube having relatively larger radius remained stable at 800 K
after 250 time steps. In contrast to $(n,n)$ SWSiNTs which are
found unstable for $n < 6$, the (3,3) SWCNT is known to be stable
and experimentally fabricated.\cite{tang,nwang} The difference in
the chemical behavior of C and Si can be traced to the difference
in their $\pi$-bonding capabilities. Si tends to utilize all of
its three valence $p$-orbitals, resulting in $sp^3$-hybridization.
In contrast, the relatively large promotion energy from
C-\emph{2s} to C-\emph{2p} orbitals explains how carbon will
activate one valence $p$-orbital at a time leading, in turn, to
$sp, sp^2, sp^3$-hybridizations in 1D, 2D and 3D structures. This
is the explanation why tubular structures of C are more stable
than those of Si.\cite{RQZhang} Moreover, since the interatomic
distance increases significantly in going from C to Si, the $\pi -
\pi$ overlap decreases accordingly, resulting in much weaker
$\pi$-bonding for Si tubes in comparison with that for carbon
tubes.

\begin{figure}
\includegraphics[scale=0.4]{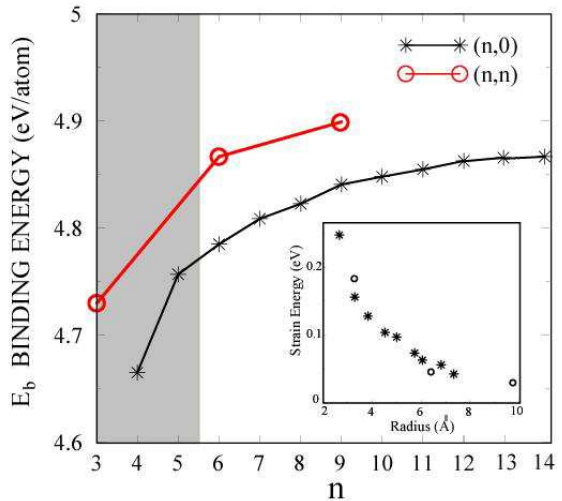} \caption{ The calculated binding energy
per atom for $(n,0)$ zigzag and $(n,n)$ armchair SWSiNTs. Dark
region indicates that tubular structures are unstable at finite
temperature. The calculated strain energies, E$_S$ per atom for
$(n,0)$ and $(n,n)$ SWSiNTs are shown by inset.}
\label{fig:binding}
\end{figure}

After the discussion of stability, we next analyze the energetics,
namely the behavior of binding energy ($E_b$) as a function of the
radius (or $n$) of the tube.  $E_b$ per atom is calculated using
the expression,
\begin{equation}\label{equ:1}
    E_b=\{E_T[SWSiNT]-N(E_T[Si])\}/N
\end{equation}

in terms of the total energy of the optimized SWSiNT having $N$ Si
atoms per unit cell, $E_T[SWSiNT]$, and the total energy of $N$,
free Si atom $E_T[Si]$. It is found that $E_b \sim$ 4.9 eV and
slightly increases as the radius $R$ (or $n$) increases for both
zigzag and armchair SWSiNTs as displayed in Fig.\ref{fig:binding}.
The energy increase with $n$ is small. According to our results
$E_b$'s of $(n,n)$ armchair SWSiNTs are $\sim 0.05$ eV larger than
those of $(n,0)$ zigzag ones because of their relatively larger
radius at a given $n$. Corresponding $E_b$ for SWCNTs is
calculated to be 9.1 eV\cite{durgun} theoretically.

Finally, the strain energy per atom is calculated relative to the
energy of honeycomb structure,

\begin{equation}\label{equ:strain}
    E_S=E_b[SWSiNT]-(E_b[Si-HC])
\end{equation}

by subtracting the binding energy (per atom) of optimized
honeycomb structure, $E_b[Si-HC]$ from the binding energy of
SWSiNT. Slight increase in strain energy is observed as the radius
$R$ or $n$ decreases. This is an expected result, since the
structure becomes more graphene-like with the increasing radius.
Calculated strain energies given by inset in Fig.\ref{fig:binding}
are also in agreement with the results obtained by Fagan \emph{et
al.}\cite{Fagan2} and Barnard \emph{et al.}\cite{Banard}
Calculated value of the strain energy of a zigzag SWSiNT is
smaller than the strain energy of a zigzag SWCNT having comparable
radius.\cite{gulseren} In classical theory of elasticity the
strain (or curvature) energy of a tubular structure is given by
the expression $E_S=\alpha / R^2$, where $\alpha$ is a function of
Young`s modulus and thickness of the tube's
wall.\cite{review,robert} Result of the present calculations in
Fig.\ref{fig:binding} gives a fair fit to the expression, $\alpha
/ R^2$ with $\alpha \sim 2.07 eV/\AA^2$.

\subsection{Mechanical Properties}

Radial flexibility is a criterion for the stability of tubular
structure. SWCNTs are known to be flexible for the deformations in
radial directions;\cite{tombler,pzhang} they can sustain severe
radial deformation transforming the circular cross section into an
elliptical one with minor ($b$) and major ($a$) axis. The radial
deformation on a bare tube of radius \emph{R} is specified in
terms of the strain associated with the pressing of the tube along
the minor axis, $\epsilon_{yy}=(b-R)/R$ and the strain associated
with the expansion of the tube along the perpendicular major axis
$\epsilon_{xx}=(a-R)/R$. Theoretical and experimental research
have shown that radially deformed tubes relax reversibly to
original circular cross section whenever the external radial force
is lifted.\cite{reversible} Moreover, radial deformation can
modify the electronic structure reversibly, that leads to a
tunable band gap engineering.\cite{reversible,kilic} For example,
a semiconducting (\emph{n},0) can be metallic under radial
deformation. Our results indicate that SWSiNTs display a behavior
different than that of SWCNTs. We performed a systematic analysis
of radial strain for (8,0) zigzag and (6,6) armchair SWSiNTs.
First, these tubes have been deformed by applying
$\epsilon_{yy}$=-0.1,-0.2 and -0.3. Then the stress (or
constraint) imposing these radial strains has been lifted and the
structure has been optimized. Contrary to situation in carbon
nanotubes, up to the applied strain $\epsilon_{yy} \leq$-0.2 the
SWSiNTs have remained in deformed state. For example, (8,0) tubes
with initial radial strain of $\epsilon_{yy}$=-0.1 and -0.2 are
relaxed to a plastic deformation corresponding to
$\epsilon_{yy}$=-0.09 and $\epsilon_{yy}$=-0.14, respectively.
Similar results have been obtained for (6,6) armchair SWSiNT with
initial radial strain of $\epsilon_{yy}$=-0.1 and -0.2. In
contrast, the tubes, which initially strained by
$\epsilon_{yy}$=-0.25 and -0.3 have relaxed to a state with
negligible residual strain. The total energy of the undeformed
SWSiNT $E_{T}^{o}$ have been found to be lower (more energetic)
than the total energy $E_{T}^{r}$($\epsilon_{yy}$) of tubes which
were relaxed upon radial deformation $-0.3 \leq \epsilon_{yy} \leq
0$. However, the energy difference $\Delta
E=E_{T}^{r}(\epsilon_{yy})-E_{T}^{o}
> 0$ is very small. The weakness of $\pi$-bonds of Si as compared
to carbon nanotubes is possibly a reason why the restoring forces
are not strong enough to derive the deformed state to relax back
to the original undeformed state. Once the applied radial
deformation gets significant ($|\epsilon_{yy}|>0.2$) the restoring
forces become strong enough to derive the relaxation towards
circular cross section. On the other hand, after a severe radial
strain that causes to a significant coupling between opposite
internal surfaces the deformed state may be more energetic
(\emph{i.e} $E_{T}^{r}(\epsilon_{yy})< E_{T}^{o}$ or it may relax
to different structures such as clusters. This situation
constitutes an important difference between Si and C single-wall
nanotubes.

Axial strength of SWSiNT, or the elastic stiffness along tube axis
is defined as the second derivative of the strain energy per atom
with respect to the axial strain $\epsilon_{zz}$, namely $\kappa
=d^{2}E_{T}/d\epsilon_{zz}^{2}$. The elastic stiffness of the
(8,0) SWSiNT along its axis is calculated to be 23 eV. This value
is significant, but smaller than that of SWCNT which is calculated
to be 52-60 eV.\cite{portal}

\subsection{Electronic Structure}

\begin{figure}
\includegraphics[scale=0.4]{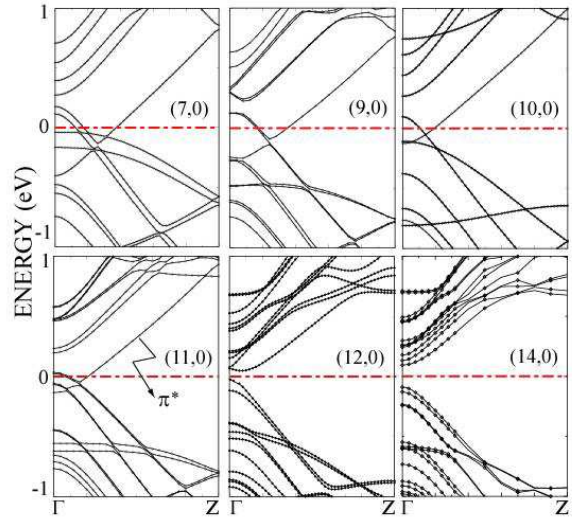} \caption{The energy bands calculated
for (7,0), (9,0), (10,0), (11,0),(12,0) and (14,0) SWSiNTs using
GGA. The lowest conduction band, or singlet $\pi^*$-band is
indicated. The zeros of energy are set at the Fermi level.}
\label{fig:allband}
\end{figure}

A systematic analysis of electronic structure indicates that
metallic zigzag SWSiNTs $6 \leq n \leq 11$ have three bands
crossing the Fermi level, but a band gap between valence and
conduction bands opens when $n \geq 12$. Similar effect has been
obtained for zigzag SWCNTs when $n \geq 7$.\cite{review,gulseren}
This metal-semiconductor transition was attributed to the energy
shift of the singlet $\pi^*$-band which is normally empty, but
becomes filled due to increased $\sigma^*-\pi^*$ hybridization at
small radius.\cite{gulseren,blase} In the present case it appears
that $\sigma^*-\pi^*$ hybridization becomes significant at
relatively larger radius. The conductance of all these infinite,
perfect tubes ($6 \leq n \leq 11$) is predicted to be equal to
$3G_o$ ($G_o=2e^2/\hbar$). Similar metallic behavior is also
obtained for armchair types namely for (6,6) and (9,9) SWSiNTs.
The conductance of ideal infinite $(n,n)$ tubes is $2G_o$, but not
$3G_o$ as in metallic $(n,0)$ zigzag tubes. Fig. \ref{fig:allband}
presents the systematic analysis of $(n,0)$ tubes for $7 \leq n
\leq 14$ and clearly shows how the singlet $\pi^*$-band gradually
raises as \emph{R} increases.

Based on LDA calculation Fagan \emph{et al.}\cite{Fagan1,Fagan2}
also found (6,6) and (6,0) SWSiNT`s metallic, but they predicted
(10,0) and (12,0) zigzag nanotubes are semiconductor with a small
band gap of 0.1 eV. The disagreement between the present one and
those of Fagan \emph{et al.}\cite{Fagan1,Fagan2} may be due to the
differences in pseudopotentials and in the approximation of
exchange correlation potential. Note that the transition of
$(n,0)$ SWSiNTs from metallic to semiconducting state through gap
opening may occur at $n$ that is smaller than predicted by the
present study as well by Fagan \emph{et al.}, if self-energy
correction are taken into account by GW method.\cite{gw}
Electronic structure analysis performed for the tubes under strain
both radially and axially showed that metallic character is not
altered but only the position of Fermi level slightly changed due
to deformation. The modification of electronic structure with
chirality may offer the possibility of fabrication of nanodevices
using SWSiNT junctions. On the other hand, SWSiNTs can be used as
metallic interconnects, since their conductance is not severely
affected by deformation.

\subsection{Interaction of SWSiNT with Atoms and Molecules}

\begin{figure}
\includegraphics[scale=0.45]{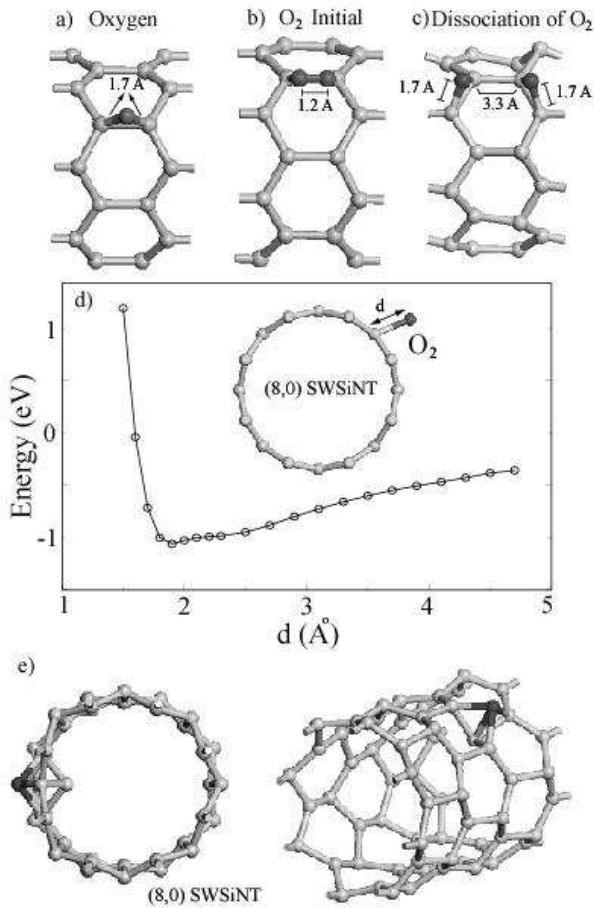}
\caption{(a) The optimized geometry of O atom adsorbed over the
axial site of (8,0) SWSiNT. (b) Initial adsorption geometry of
$O_2$ over the axial site. (c) Dissociation of $O_2$ after
relaxation and formation of Si-O-Si bridge bands over the adjacent
zigzag Si-Si bonds. (d) Variation of interaction energy as a
function of the distance $d$ of $O_2$ molecule from the parallel
axial Si-Si band. (e) Optimized geometry of a single Si atom
adsorbed on the top site (indicated by a dark ball).}
\label{fig:oxidation}
\end{figure}

The interaction of Si nanotubes with oxygen atom and oxygen
molecule is extremely important for technological applications.
The adsorption of oxygen atom is studied by placing it initially
above a Si-Si bond parallel to the axis of a (8,0) SWSiNTs. The
optimized structure  shown in Fig. \ref{fig:oxidation}a. has a
very strong chemical bonding between O and SWSiNT with $E_b$=8.1
eV and the nearest Si-O distance 1.7 $\AA$. The resulting geometry
showed that SWSiNT is slightly distorted upon O adsorption.

The interaction between O$_2$ and SWSiNT has been revealed by
calculating the binding energy as a function of the separation $d$
from the axial Si-Si bond of the tube as shown in Fig.
\ref{fig:oxidation}(b). $O_{2}$ molecule is kept unrelaxed and
taken parallel to the Si-Si bond. The calculated energy versus
distance curve $E(d)$ in Fig. \ref{fig:oxidation}(d) shows that
O$_2$ can be attracted to the tube, but there is no physisorption
state as in O$_2$+SWCNT.\cite{sefa} The minimum of $E(d)$ occurs
at 1.9 $\AA$. Upon relaxation of the tube and O$_2$ near this
minimum, the molecule has been dissociated to form two Si-O-Si
bridge bonds over the zigzag Si-Si bonds, and concomitantly SWSiNT
has been distorted locally as illustrated in
\ref{fig:oxidation}(c). The distances between nearest Si-O and O-O
are 1.7 and 3.3 $\AA$, respectively. We repeated the structure
relaxation by initially placing O$_2$ at a larger distance $d=2.5
\AA$ from the surface of the tube and we obtained the same
dissociated state. Our results indicate that there will be a
strong interaction between Si nanotube and oxygen molecule in open
air applications.

The SWSiNT surface is found to be reactive against Si, H and Au
atoms. Si atom attached to the top site is bound by E$_b \sim 5$
eV. One Si atom of the tube is plunged inside the tube and a small
cluster is formed at the surface (see Fig.
\ref{fig:oxidation}(e)). The chemisorption energy of H and Au
atoms is strong and found to be 4.4 eV and 3.4 eV, respectively.

\section{Stabilization of Silicon Nanotubes by Transition Metal Atom Doping }

\begin{figure}
\includegraphics[scale=0.4]{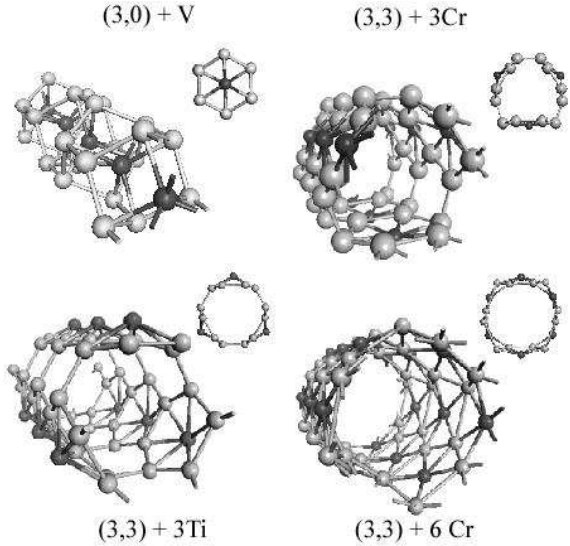}
\caption{Perspective and top (cross section) view of the fully
relaxed V-doped (3,0), Ti and Cr doped (3,3) SWSiNTs. Our results
revealed that small radius SWSiNTs can be stabilized by doping
transition metal element.} \label{fig:doping}
\end{figure}

Recently, Singh \textit{et al.}~\cite{Singh} showed that
Si-clusters and Si-tubular structures formed by top-to-top
stacking of Si-hexagons can be stabilized by the implementation of
TM atoms inside these structures. Those structures are not only
stabilized, but also acquired magnetic properties. Earlier, TM
atoms are shown to form rather strong bonds with the carbon atoms
on the surface of SWCNTs.\cite{durgun} Motivated by the work of
Singh \textit{et al.}\cite{Singh}, we investigated whether (3,0)
SWSiNT can be stabilized in the same manner. The (3,0) tube has
radius $R\sim2.4\AA$ in which interaction between the atoms
located at the opposite walls of the tube as well as excess strain
on the Si-Si bonds are the prime causes of structural instability
even at T=0~K. On the other hand, the radius of (3,0) is
comparable with the sum of ionic radii of V and Si, \emph{i.e}
$R_{V}+R_{Si}=2.27\AA$, and hence V atoms can easily be
accommodated inside the tube. We considered a (3,0) SWSiNT, which
has V atoms implemented inside and periodically arranged along the
tube axis. Because of supercell geometry used in the calculations
both chains of V atoms (V-LC) and (3,0) tube have common lattice
parameter (See Fig.\ref{fig:doping}). The optimized structure,
that consists of planar hexagons are stacked with V-LC passing
through their centers, has been found to be stable. The energy of
the V-stabilized structures is lowered by 12.9 eV relative to the
energies of the V-LC and Si-tube without V-LC in it but having the
same atomic structures as Si tube with V-LC. Spin-relaxed
calculations resulted in zero magnetic moment $\mu=0$. The
\emph{p-d} hybridization between Si and V atomic orbitals is the
cause of stability and lowering the total energy.

The radius of the unrelaxed (3,3) SWSiNT of 3.7 $\AA$ is too large
and may not be suitable for its stabilization through the
implementation of an atomic chain. Perhaps, it may be better
suited to accommodate a small clusters of atoms. We considered the
possibility whether the (3,3) SWSiNT is stabilized by TM atoms
adsorbed on the external surface of the tube. To this end, we
studied Ti and Cr atoms adsorbed on the hollow sites (\emph{i.e}
above the hexagons formed by Si atoms). As shown in
Fig.~\ref{fig:doping} we examined 3 different (3,3)+TM structures;
namely 3 Ti-LC, 3 Cr-LC and 6 Cr-LC are adsorbed on the (3,3)
SWSiNT surface. Spin-relaxed GGA calculation are carried out to
optimize the geometric structure. The external absorption 3 Ti-LC
(or 3 Ti atom per unit cell of (3,3) tube) prevented the tube from
collapsing into a cluster, but the circular cross section changed
to a polygonal one. The ground state has been predicted to be
non-magnetic with $\mu=0$. The external adsorption of 3 Cr-LC also
resulted in polygonal cross section, but ferromagnetic ground
state with net magnetic moment $\mu=9.7 \mu_B$ (Bohr magneton).
The circular cross section is maintained by the adsorption of 6
Cr-LC. This latter structure has also ferromagnetic ground state
with $\mu=17.2 \mu_B$.

\begin{figure}
\includegraphics[scale=0.4]{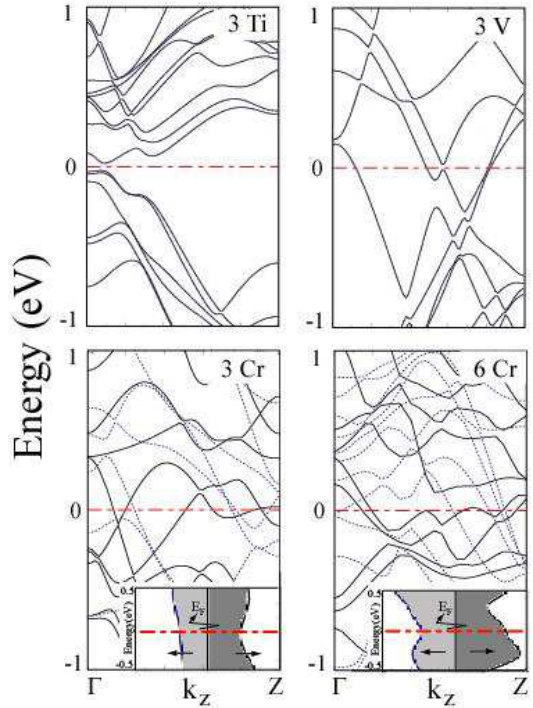}
\caption{The energy band structures of (3,0)+V, (3,3)+3Ti,
(3,3)+3Cr and (3,3)+6Cr calculated by using SCF spin polarized GGA
method. Solid lines and dotted lines are are for majority and
minority states, respectively. The inset shows the density of
majority (dark) and minority (light) spin states at the Fermi
level of (3,3)+3Cr and (3,3)+6 Cr structure.} \label{fig:tm}
\end{figure}

Calculated energy band structure of (3,0)+V, (3,3)+Ti, (3,3)+3Cr
and (3,3)+6Cr are presented in Fig. \ref{fig:tm}. The (3,0)+V
structure is a metal. Six bands crossing the Fermi level yields
quantum ballistic conductance of G=6$G_0$. The partial density of
states indicates that V-\emph{3d} and Si-\emph{3p} orbital
character dominate  the states at the Fermi level. The (3,3)+3Ti
structure is a semiconductor with a very narrow band gap. In the
case of (3,3)+3Cr and (3,3)+6Cr several majority (spin-up
$\uparrow$) and minority (spin-down $\downarrow$) bands are
crossing the Fermi level. Hence both structures are metals with
finite density of majority $D(E=E_{F},\uparrow)$ and minority
$D(E=E_{F},\downarrow)$ spin states at E$_F$. However,
$D(E=E_{F},\uparrow)-D(E=E_{F},\downarrow)$ is significant. These
properties, which are also depends on the decoration of the tubes,
can be used in nanospintronic device applications. Much recently
Dumitrica \emph{et al.}\cite{dumitrica} have investigated the
stabilization (3,0) zigzag and (2,2) armchair SWSiNTs by various
atoms (Zr, Sc, Ti, Cr, Fe, Ni, Be, And Co) axially placed inside
the tube. However, they consider neither the magnetic ground state
due to specific TM atoms, nor the stabilization of tubes having
relatively larger radius.

\section{other tubular structures of silicon}

\begin{figure}
\includegraphics[scale=0.4]{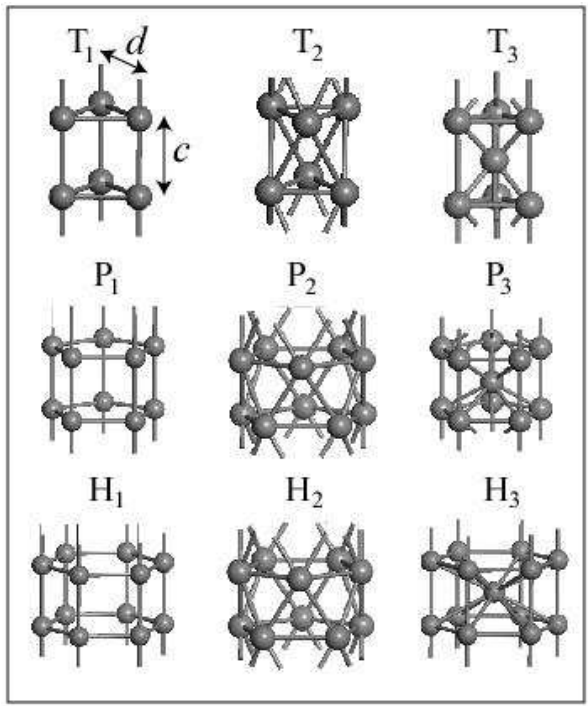}
\caption{Some tubular structures of Si. The tubes are formed by
stacking of Si polygons; the labels T, P, and H stand for
triangular, pentagonal, and hexagonal configurations,
respectively. T$_1$, P$_1$, and H$_1$ tubes have top-to-top
arrangement of layers. In T$_2$, P$_2$, and H$_2$ the layers are
staggered. T$_3$, P$_3$, and H$_3$ structures have extra atoms
centered in between the layers.} \label{fig:tube}
\end{figure}

We now consider different tubular structures which are formed by
stacking of atomic polygons as presented in Fig. \ref{fig:tube}.
Three tubular structures given in the first row, \emph{i.e.}
T$_{1}$, T$_{2}$ T$_{3}$ are made of triangles of silicon atoms,
which are stacked along the axis of the tube. In T$_{1}$,
triangles are identical and placed in top-to-top (or eclipsed)
position; in T$_{2}$ the triangles are staggered; T$_{3}$
structure is constructed by insertion of a Si-LC into the T$_{1}$
structure as such that the chain atoms are centered in between the
layers. The same convention is followed in labelling the tubular
silicon nanowires with pentagonal (P$_{1}$, P$_{2}$, P$_{3}$) and
hexagonal (H$_{1}$, H$_{2}$, H$_{3}$) cross-sections.

\begin{table}[h]
\caption{Structural and conductance properties of silicon tubular
structures that are found stable. $E_b$ is the binding energy per
atom, $c$ is the unit cell length of the periodic structure, $d$
is the in-plane bond lengths of the polygonal atomic layers.
Equilibrium conductance values are given by $G$ in units of
conductance quantum, $G_o = 2e^2/h$ }\label{tab:tube}
\begin{center}
\begin{tabular}{ccccc}
     \hline \hline

  Structure~~~   & ~~~$E_b$ (eV)~~~ & ~~~$c$ (\AA)~~~ & ~~~$d$ (\AA)~~~ & ~~~$G$ ($2e^2/h$) \\
     \hline %

     T$_1$& 4.62 & 2.37 & 2.38 & 6 \\
     T$_2$& -- & -- & -- & -- \\
%
%     T3& 4.63 & 4.07 & 2.66;2.44 & 0.14 eV \\
      T$_3$& -- & -- & -- & -- \\
     P$_1$& 4.79 & 2.37 & 2.37 & 10 \\
     P$_2$& 4.68 & 2.70 & 2.46 & 10 \\
     P$_3$& 4.66 & 4.10 & 2.58 & 6 \\
     H$_1$& 4.77 & 2.29 & 2.37 & 6 \\
     H$_2$& 4.74 & 2.52 & 2.42 & 9 \\
     H$_3$& -- & -- & -- & -- \\
%
%
%     H4& 4.x & 3.x & 4.x & 6 \\
     \hline \hline%

\end{tabular}
\end{center}

\end{table}

\begin{figure}
\includegraphics[scale=0.35]{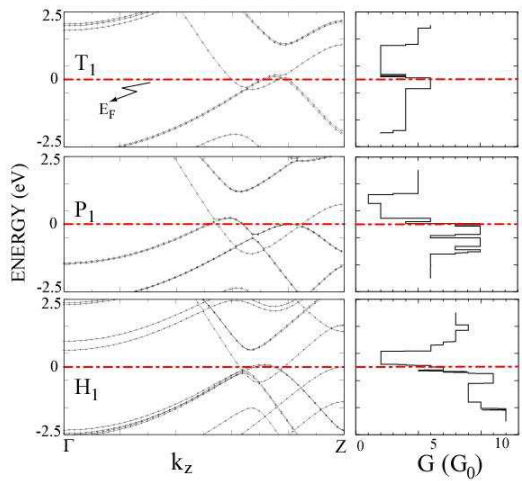}
\caption{Energy band structure of Si tubular structures formed by
top-to-top stacking of triangular (T$_1$), pentagonal (P$_1$), and
hexagonal (H$_1$) polygons. In the right panels the corresponding
equilibrium conductance in $G_o = 2e^2/h$ curves are plotted. The
Fermi levels are set to zero.} \label{fig:tube-band}
\end{figure}

Starting with those geometries, structural optimizations in large
supercells yielded T$_2$, T$_3$, and H$_3$ as unstable, which
deformed into clusters. The structural parameters, the binding
energies, and equilibrium conductance values of stable tubular
structures are summarized in Table \ref{tab:tube}. All the stable
silicon tubes have comparable binding energy values, with P$_1$
structure being slightly more favorable energetically. Also within
each set of tubular structures, the top-to-top arrangement
provides the highest cohesive energy. In Fig. \ref{fig:tube-band}
the energy band structure and equilibrium conductance plots of
T$_1$, P$_1$, and H$_1$ structures are presented. All of them are
found metallic with calculated ideal conductance values of 6$G_0$,
10$G_0$, and 6$G_0$, respectively. A common feature noticed in the
energy band structures of T$_1$, P$_1$, and H$_1$ tubes is that
there are almost filled bands available in close vicinity of the
Fermi levels which may lead to drastic conductance variations due
to small structural perturbations in the tubes or small bias
voltage.

Earlier, the pentagonal nanowires of Si and also those of several
metals such as Na, Al, Cu, Pb, Au, Fe, Ni, and Xe were
investigated by Sen \textit{et al.} \cite{sen} using similar
calculation methods. The results for a specific structure, P$_3$,
which is common in both studies, is in agreement. The stability of
the P$_3$ is further strengthened by the finite-temperature
\textit{ab-initio} molecular dynamics calculation carried out in
the present study. Recently, Bai \emph{et al.} \cite{bai} have
studied the stability of infinite and finite S$_1$ (top-to-top
square), P$_1$ and H$_1$ structures by performing \emph{ab-initio}
calculations using different methods including pseudopotential
plane-waves and classical molecular dynamics calculations at
2000K. Their stability analysis for P$_1$ and H$_1$ are in
agrement with present results. In addition to P$_1$ and H$_1$, the
present study deals with T$_1$ and also staggered ones.

\section{single wall nanotubes of III-V compounds}

\begin{figure}
\includegraphics[scale=0.5]{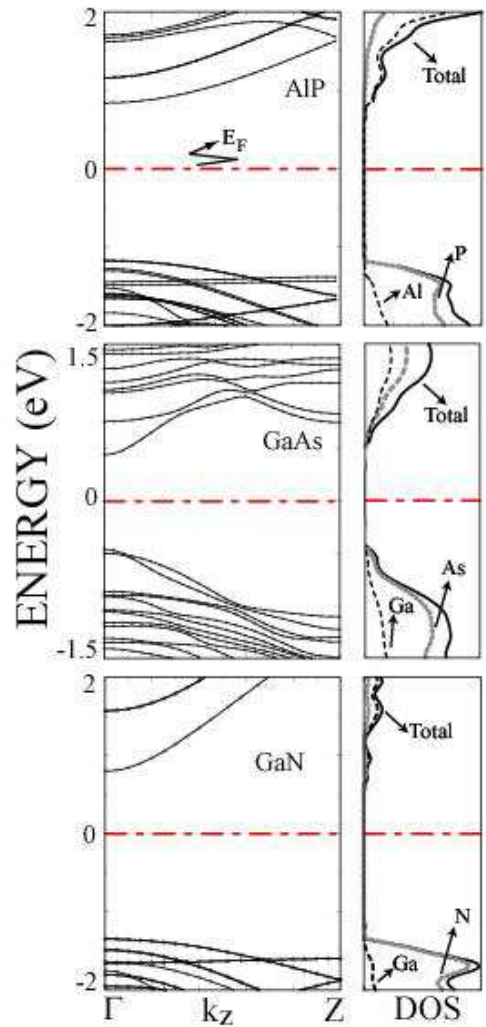}
\caption{Energy band structures (left panels),total density of
states (TDOS) and partial density of states (PDOS) on atoms (right
panels) of (8,0) zigzag SW(AlP)NT, SW(GaAs)NT, and SW(GaN)NT.
Anion (Al, Ga) and cation (P, As, N) contributions to TDOS are
shown by dashed and light-continuous lines. Zero of the energy is
set to the Fermi energy.} \label{fig:alp_gaas}
\end{figure}

Motivated by interesting properties of SW(BN)NT and
opto-electronic and field emitting properties of GaN and AlN
tubular forms \cite{tondare,zhao} we choose (8,0) AlP, GaAs and
GaN single-wall nanotubes as prototype to investigate the
stability and electronic properties of III-V compound nanotubes.
Even if the single-wall nanotubes of these compounds have not been
sythesized yet, the predictions of present work is essential for
further efforts to achieve it. The initial bond lengths are chosen
as the distance between nearest cation and anion atoms in bulk
structure. After relaxation of all atomic positions, as well as
lattice constant $c$, the tubular structures remained stable. The
\emph{ab-initio} MD calculations also showed that SW(AlP)NT
remained stable at room temperature after 250 time steps. $E_b$ is
calculated to be 9.6 eV per AlP basis. The radius of the tube is
5.2 $\AA$. The structure is not a perfect tube but the hexagons on
the surface are buckled. The nearest Al-P distance is 2.3$\AA$,
and second nearest neighbor distance $i.e$, nearest P-P and Al-Al
distances are 3.9$\AA$ and 3.8$\AA$, respectively. The energy band
and TDOS analysis in Fig. \ref{fig:alp_gaas} points out that (8,0)
SW(AlP)NT is a semiconductor (insulator) with a band gap of 2.0
eV.

Initial tubular structure of (8,0) SW(GaAs)NT is maintained after
geometry optimization at T=0 K. Similar to SW(AlP)NT, hexagons are
buckled. $E_b$ is calculated to be 7.7 eV per GaAs and the radius
is 4.8 $\AA$. The nearest Ga-As distance is 2.4$\AA$, and nearest
Ga-Ga and As-As distances are 3.9$\AA$ and 4.1$\AA$, respectively.
The (8,0) SW(GaAs)NT is also a semiconductor (insulator) with a
band gap of 0.9 eV.

We place a special emphasis on GaN
nanotubes.\cite{goldberger,wlee} which are successfully
synthesized by an epitaxial casting method where ZnO nanowires are
initially used as templates. GaN nanotubes produced this way have
diameter of 300 $\AA$ and minimum wall thickness of 50 $\AA$. They
are semiconducting and hence they would be a possible candidate
for optoelectronic applications. Whether single-wall  GaN tube of
smaller diameter ($2R \sim 10 \AA$) can be stable and can exhibit
technologically interesting electronic properties is important to
know. We again took (8,0) SW(GaN)NT as prototype for the sake of
consistency. Stable tubular geometry is obtained by both geometry
optimization at T=0 K and \emph{ab-initio} MD analysis at T=800 K.
Upon relaxation atoms on the surface are buckled. $E_b$ is
calculated to be 11.5 eV per GaN and the radius is 4.1 $\AA$. The
nearest Ga-N distance is 1.8$\AA$, and nearest Ga-Ga and N-N
distances are 3.1$\AA$ and 3.2$\AA$, respectively. We found that
the (8,0) SW(GaN)NT is a semiconductor (insulator) with a band gap
of 2.2 eV. Previous first-principles study performed by Lee et
al.\cite{smlee} by using LDA method predicted the similar band gap
for SW(GaN)NT. Here, we examine also whether SW(GaN)NT is radially
elastic. To this end we started with the elliptically deformed
nanotube under $\epsilon_{yy}=-0.1$, and let it relax in the
absence of radial forces. Similar to the Si nanotube, SW(GaN)NT is
found to be radially soft

For the sake of comparison, the binding energy and bond distance
of bulk crystal, honeycomb structure, and (8,0) tubular structures
of Si, AlP, GaN, and GaAs are presented in Table \ref{tab:table}.
In these structures the covalent or (covalent-ionic) mixed bonds
have different orbital combinations. While bulk crystals are
tetrahedrally coordinated and have bonds formed by $sp^3$-hybrid
orbitals, in honeycomb and tubular structures bonding through
$sp^2$-hybrid orbitals dominates the cohesion.

\begin{table}
\begin{center}
\begin{tabular}{c|c|c|c|c}


      &Si&AlP&GaN&GaAs\\
     \hline%
     &B~ H~ T & B~ H~ T & B~ H~ T & B~ H~ T\\
     \hline\hline
     E$_b$ (eV) & 5.4  4.9  4.8 & 10.4 9.6 9.6 & 12.4 11.8 11.5 &
     8.3 7.2 7.7\\
     d ($\AA$) & 2.3 2.3 2.3 & 2.4 2.2 2.3 & 2.0 1.8 1.8 & 2.4 2.2
     2.4
\end{tabular}
\end{center}
\caption{Calculated binding energies $E_b$ and bond distances
\emph{d} in $\AA$ of various structures of Si, AlP, GaAs and GaN.
B:3D bulk crystal, H:2D honeycomb structure, T:(8,0) single wall
nanotube structure. The units of binding energy is eV/atom for Si,
and eV/basis for III-V compounds.}\label{tab:table}
\end{table}

\section{conclusion}

In this paper, we analyzed the stability of Si and III-V compound,
single-wall nanotubes and calculated their optimized atomic
structure and energy band structure. Si as well as III-V compounds
can form stable 2D honeycomb structure, which is precursor of
nanotubes. The energy necessary to roll these honeycomb structures
over a cylinder of radius $R$ to make a perfect nanotube is
however small as compared to those in carbon nanotubes. We found
that Si single-wall nanotubes with small radius are unstable and
are clustered either at T=0 K or at finite temperatures. For
example, while (3,0) is unstable even at T=0 K, (4,0) and (5,0)
lose their tubular character and tend to form  cluster at T=500 K.
Stable $(n,0)$ zigzag SWSiNTs are metallic for $6 \leq n \leq 11$,
but become semiconducting for $n \leq 12$. The metallicity of
small radius $(n,0)$ tubes is a typical curvature effect and is
resulted from the dipping of the singlet $\pi^*$-band into the
valence band at small radius. Stable $(n,n)$ armchair SWSiNTs
 ($n$=6,9)are metallic. Our study on radially deformed (8,0) and (6,6)
SWSiNTs demonstrated that these nanotubes are radially "soft", and
hence are devoid of strong restoring force that maintains radial
elasticity. The radial softness of Si tubes is a behavior which
distinguishes them from carbon nanotubes. In contrast to that
axial stiffness the Si nanotube has ben found to be high. We
predicted that oxygen molecule adsorbed on the Si-Si bonds
dissociates. A strong interaction between O/O$_2$ and SWSiNT
appears to be serious in future processes involving Si tubes.
Adatoms like Si, Au and H can also form strong chemisorption bonds
with the atoms on the surface of SWSiNT. We showed that unstable,
small radius SWSiNTs can be stabilized through the implementation
or external adsorption of $3d$ transition metal atoms. In
particular, the decoration of the tube surface by the external
adsorption of transition metal atoms can lead magnetic properties
which may find potential technological applications. Small radius
tubular structures different than those based on honeycomb
structure have been identified. Finally, we found III-V compound
(8,0) nanotubes (AlP, GaAs, and GaN) stable at least at room
temperature and they are semiconductor with band gap ranging from
0.9 eV to 2.2 eV. In contrast to small radius metallic Si
nanotubes, (8,0) compound nanotubes are semiconductor. The band
gap increases with decreasing row number of elements. Even though
not all the structures treated in this study have not been
realized experimentally yet, the predictions obtained from the
present first-principles calculations are expected to be essential
for further research in this field.

\begin{acknowledgments}
Part of computations have been carried out at ULAK-BIM Computer
Center. SC acknowledges partial financial support from Academy of
Science of Turkey.
\end{acknowledgments}

\newpage

\end{document}